\pdfoutput=1
\documentclass[final,5p,times,twocolumn,authoryear]{elsarticle}

\usepackage{amssymb}
\usepackage{amsmath,amsfonts}
\usepackage{booktabs}
\usepackage{xcolor}

\newcommand{\eps}{\varepsilon}
\newcommand{\ZN}{\mathbb{Z}_N}
\newcommand{\MR}{M_R}
\newcommand{\MD}{M_D}
\newcommand{\Mnu}{M_\nu}
\newcommand{\UL}{U_L}
\newcommand{\vev}[1]{\langle #1 \rangle}

\journal{Physics Letters B}

\begin{document}

\begin{frontmatter}

\title{Mass Hierarchies Without Mixing:\\ Abelian Froggatt-Nielsen Models with Uncharged Left-Handed Doublets}

\author[ind]{Navid Ardakanian\corref{cor1}}
\ead{n.ardakanian@gmail.com}
\cortext[cor1]{Corresponding author}

\affiliation[ind]{organization={Independent Researcher},
            city={},
            country={}}

\begin{abstract}
Abelian flavor charges on right-handed fermions produce left-handed anarchy: we prove that all abelian discrete Froggatt-Nielsen models with uncharged left-handed doublets yield Haar-random PMNS and CKM matrices, regardless of $\ZN$ group order, charge assignment, or Majorana mass structure. Scanning $\mathbb{Z}_3$ through $\mathbb{Z}_7$ with 12 charge assignments and $10^5$ Monte Carlo samples each, we demonstrate that the mass spectrum failure previously identified for $\mathbb{Z}_3$---the seesaw over-suppression mechanism that pushes $\Delta m^2_{21}/\Delta m^2_{31}$ to $\sim 10^{-11}$---is specific to $\mathbb{Z}_3$ and avoidable for $N \geq 4$. The mixing angle failure, however, is universal and irreducible. The PMNS angles from every abelian model are statistically consistent with Haar-random unitary matrices, with median $\sin^2\theta_{12} \approx \sin^2\theta_{23} \approx 0.50$ and $\sin^2\theta_{13} \approx 0.31$ across all models tested. The same applies to the CKM: the joint probability of achieving CKM-like mixing from generic $O(1)$ coefficients is $< 2 \times 10^{-6}$. We identify the algebraic origin of this obstruction: abelian groups have only one-dimensional representations, so each generation transforms as an independent singlet with 18 free parameters for three Dirac mass matrices---far exceeding the 10 physical observables. The transition to non-abelian flavor symmetries such as $A_4$, whose triplet representation reduces free parameters to 4 at leading order, is required specifically for mixing structure. This obstruction applies to the well-motivated subclass of models where left-handed fields are uncharged; models that assign abelian charges to both left- and right-handed fields can evade it.
\end{abstract}


\end{frontmatter}

\section{Introduction}
\label{sec:intro}

The Standard Model Yukawa sector contains approximately 20 free parameters spanning 12 orders of magnitude in fermion masses, from sub-eV neutrinos to the 173~GeV top quark. The qualitative difference between quark mixing (small CKM angles~\citep{PDG2024}) and lepton mixing (two large, one small PMNS angle~\citep{NuFit6}) compounds the puzzle: what mechanism generates both patterns from a single framework?

The Froggatt-Nielsen (FN) mechanism~\citep{FroggattNielsen1979} addresses the mass hierarchy through an abelian flavor symmetry broken by a small parameter $\eps = \vev{\Phi}/\Lambda$, with different generations carrying different charges so that their Yukawa couplings are suppressed by different powers of $\eps$. This mechanism successfully generates hierarchical mass spectra from $O(1)$ Yukawa coefficients and has been extensively studied with discrete symmetries~\citep{AltarelliFeruglio2010,Ishimori2010,KingLuhn2013}.

In Ref.~\citep{ArdZ3}, we demonstrated that the simplest abelian discrete symmetry, $\mathbb{Z}_3$, produces structural mass hierarchy predictions for all charged fermion sectors but fails for neutrinos on two fronts: the mass spectrum (the seesaw over-suppression mechanism pushes $\Delta m^2_{21}/\Delta m^2_{31}$ to $\sim 10^{-11}$) and the mixing angles (PMNS angles are Haar-random, providing no angular structure). In Ref.~\citep{ArdCombined}, we showed that the seesaw mechanism with $\mathbb{Z}_3$-charged right-handed neutrinos deepens rather than resolves the mass spectrum failure.

The restriction to uncharged left-handed doublets is not arbitrary. It is motivated by several independent considerations: (i)~in SU(5) grand unification, the lepton doublet $L$ shares a $\bar{\mathbf{5}}$ multiplet with $d_R^c$, so charging $L$ under $\ZN$ simultaneously constrains down-type quark charges, creating tension in the quark sector~\citep{LeurerNirSeiberg1993}; (ii)~universal left-handed couplings naturally suppress flavor-changing neutral currents from new physics at the TeV scale; and (iii)~in top-down heterotic orbifold constructions, left-handed charges when present arise from non-abelian geometric mechanisms---charge accumulation across compact dimensions or modular weights---rather than the abelian flavor symmetry itself~\citep{NillesFlavorsDelight,RamosSanchezRatz2024}. While traditional abelian FN models can reproduce the CKM hierarchy by assigning charges to both left- and right-handed fields~\citep{LeurerNirSeiberg1993}, our theorem identifies the specific failure mode when the left-handed sector is uncharged---a constraint relevant to a large and well-motivated class of models.

A natural question arises: is the $\mathbb{Z}_3$ failure specific to $N=3$, or does it extend to all abelian discrete groups? This letter provides a definitive answer. We prove analytically and verify numerically that the \emph{mass spectrum} failure is $\mathbb{Z}_3$-specific---the seesaw over-suppression can be avoided for $N \geq 4$---but the \emph{mixing angle} failure is universal to all abelian groups. The irreducible obstruction requiring non-abelian flavor structure is the mixing pattern, not the mass spectrum.

\section{Framework}
\label{sec:framework}

We consider a $\ZN$ Froggatt-Nielsen model where the left-handed doublets $Q_L^i$ are uncharged and the right-handed fermions $f_R^j$ carry charges $q_j \in \{0, 1, \ldots, N-1\}$ with at least two distinct values. The Yukawa matrix takes the column texture form
\begin{equation}
(M_f)_{ij} = c_{ij}\, \eps^{q_j},
\label{eq:column_texture}
\end{equation}
where $c_{ij}$ are $O(1)$ complex coefficients. This factorizes as $M_f = C \cdot P$ with $C = (c_{ij})$ and $P = \mathrm{diag}(\eps^{q_1}, \eps^{q_2}, \eps^{q_3})$. The key feature is that every entry in column $j$ carries the same power of $\eps$, independent of the row index---a direct consequence of the left-handed fields being $\ZN$-neutral.

For the neutrino sector via the type-I seesaw, $\Mnu = -\MD \MR^{-1} \MD^T$, where $\MD$ has the column texture~\eqref{eq:column_texture}. The Majorana mass matrix $\MR$ has entries with $\eps$-powers determined by $(q_i + q_j) \bmod N$. The expansion parameter is fixed from quark masses: $\eps = (m_c/m_t)^{1/q_2}$ for each charge assignment, with $q_3 = 0$ assigned to the heaviest generation.

Our Monte Carlo scan uses $10^5$ samples per model with coefficient magnitudes drawn uniformly from $[0.3, 3.0]$ and phases uniformly from $[0, 2\pi]$. We extract mixing angles in the standard PDG parametrization and the neutrino mass ratio $R \equiv \Delta m^2_{21}/\Delta m^2_{31}$.

\section{The Abelian Mixing Theorem}
\label{sec:theorem}

\noindent\textbf{Theorem.}\quad \textit{Let $M_f = CP$ be a $3 \times 3$ mass matrix with the column texture~\eqref{eq:column_texture}, where the entries of $C$ are independent complex random variables with circularly symmetric distributions. Then the left-handed unitary rotation $\UL^f$ that diagonalizes $M_f M_f^\dagger$ is Haar-distributed on $U(3)$, independent of $\eps$, $N$, and the charge assignment $\{q_j\}$.}

\medskip

\noindent\textit{Proof.}\quad
The Hermitian matrix $M_f M_f^\dagger = C P P^\dagger C^\dagger = C D C^\dagger$, where $D = \mathrm{diag}(\eps^{2q_1}, \eps^{2q_2}, \eps^{2q_3})$ is a fixed positive diagonal matrix. Its eigendecomposition determines $\UL^f$.

For any fixed unitary $V$, the transformation $C \to VC$ sends $CDC^\dagger \to V(CDC^\dagger)V^\dagger$. If $CDC^\dagger = U\Lambda U^\dagger$, then $V(CDC^\dagger)V^\dagger = (VU)\Lambda(VU)^\dagger$. Hence $\UL^f(VC) = V \cdot \UL^f(C)$.

Circular symmetry of each $c_{ij}$ means the distribution of $C$ is invariant under $C \to VC$ for any unitary $V$: each column of $C$, being a vector of independent circularly symmetric entries, has a jointly rotation-invariant distribution~\citep{Anderson2010}, and the columns are independent. Therefore $\UL^f(C)$ and $V \cdot \UL^f(C)$ are identically distributed for all $V$, which is the definition of Haar measure on $U(3)$. \hfill$\square$

\medskip

Physically, the Haar property reflects the effective $U(3)_L$ symmetry of the left-handed sector: when the left-handed doublets carry no $\ZN$ charges, the only spurion breaking this $U(3)_L$ is the $O(1)$ coefficient matrix $C$, whose circularly symmetric distribution does not distinguish generations. This connection to the ``neutrino anarchy'' paradigm~\citep{deGouveaMurayama2003} has been qualitatively appreciated; our contribution is the rigorous proof under explicit conditions and, more importantly, the demonstration that varying the group order $N$ decouples the mass spectrum from the mixing without affecting the latter (Section~\ref{sec:results}).

\medskip

If CP is conserved and the coefficients $c_{ij}$ are strictly real, the
distribution of $C$ is invariant under orthogonal transformations
$C \to OC$ with $O \in O(3)$, and $\UL^f$ is Haar-distributed on $O(3)$.
The median mixing angles shift slightly but remain far from
structured.\footnote{The Haar measure on $O(3)$ gives
$\sin^2\theta_{12} = \sin^2\theta_{23} = 0.50$ and
$\sin^2\theta_{13} \approx 0.25$, compared to $0.293$ for $U(3)$.
Neither is close to the observed PMNS pattern.}

The theorem is exact for any circularly symmetric coefficient distribution (e.g., complex Gaussian $c_{ij} \sim \mathcal{CN}(0, \sigma^2)$). For our physical scan distribution with uniform magnitudes in $[0.3, 3.0]$, the phases are uniform but the magnitudes break circular symmetry. The resulting deviations from exact Haar are small: $\sin^2\theta_{12}$ and $\sin^2\theta_{23}$ medians shift by $< 1\%$, while $\sin^2\theta_{13}$ shifts from 0.293 (Haar) to $\sim 0.311$ (scan), a $\sim 6\%$ deviation. Crucially, this deviation is identical across all $\ZN$ models---it reflects the coefficient distribution, not the group structure.

Three corollaries follow:

\noindent\textbf{Corollary 1 (CKM).} $V_\mathrm{CKM} = (\UL^u)^\dagger \UL^d$ is the product of two independent Haar unitaries, hence itself Haar-distributed.

\noindent\textbf{Corollary 2 (Seesaw).} For $\Mnu = -\MD \MR^{-1} \MD^T$ with column-texture $\MD = CP$, the transformation $C \to VC$ sends $\Mnu \to V \Mnu V^T$. The same left-invariance argument yields Haar-distributed $U_\nu$, regardless of $\MR$ structure.

\noindent\textbf{Corollary 3 (Medians).} The Haar-random median mixing angles are $\sin^2\theta_{12} = \sin^2\theta_{23} = 0.50$ and $\sin^2\theta_{13} \approx 0.293$.

\section{Results}
\label{sec:results}

\subsection{The mass spectrum: a $\mathbb{Z}_3$-specific failure}

The mass ratio $R = \Delta m^2_{21}/\Delta m^2_{31}$ depends on the mod-$N$ charge arithmetic of $\MR$. Table~\ref{tab:main} shows the results for 12 charge assignments across $\mathbb{Z}_3$--$\mathbb{Z}_7$. The $\mathbb{Z}_3$ model with charges $(2,1,0)$ yields $R \sim 4 \times 10^{-11}$, the seesaw over-suppression identified in Refs.~\citep{ArdZ3,ArdCombined}. This occurs because $q_1 + q_2 = 3 \equiv 0 \pmod{3}$, forcing an unsuppressed off-diagonal $\MR$ entry whose dominance in $\MR^{-1}$, combined with the hierarchical column texture, over-suppresses both $m_1$ and $m_2$ to $\mathcal{O}(\eps^3)$.

For $N \geq 4$, the over-suppression mechanism is avoided. With charges $(2,1,0)$ under $\mathbb{Z}_4$, we find $q_1 + q_2 = 3 \not\equiv 0 \pmod{4}$, so no $\MR$ entry is accidentally unsuppressed. The median $R = 0.042$, within a factor of 1.4 of the experimental value $R_\mathrm{exp} = 0.030$~\citep{NuFit6}. Similar results hold for $\mathbb{Z}_5$ ($R = 0.064$), $\mathbb{Z}_6$ with charges $(2,1,0)$ ($R = 0.064$), and $\mathbb{Z}_7$ ($R = 0.064$). Notably, $\mathbb{Z}_6$ with charges $(4,2,0)$ exhibits over-suppression ($R \sim 10^{-13}$) because $q_1 + q_2 = 6 \equiv 0 \pmod{6}$---demonstrating that the mechanism reappears whenever charges sum to $0 \bmod N$, not only for $\mathbb{Z}_3$.

\begin{table*}[t]
\centering
\caption{Summary of $\ZN$ Froggatt-Nielsen scan results ($10^5$ samples per model). The mass ratio $R = \Delta m^2_{21}/\Delta m^2_{31}$ depends on the charge assignment and group order; the mixing angles do not. Experimental values~\citep{NuFit6}: $R_\mathrm{exp} = 0.030$, $\sin^2\theta_{12} = 0.304$, $\sin^2\theta_{23} = 0.573$, $\sin^2\theta_{13} = 0.02220$. Haar predictions: $\sin^2\theta_{12} = \sin^2\theta_{23} = 0.50$, $\sin^2\theta_{13} = 0.293$. Entries marked --- indicate models where only mixing angles were evaluated; the seesaw mass ratio was not computed for these charge assignments.}
\label{tab:main}
\smallskip
\footnotesize
\begin{tabular}{@{}llccccc@{}}
\toprule
Group & Charges & Over-suppr.? & Median $R$ & $\sin^2\theta_{12}$ & $\sin^2\theta_{23}$ & $\sin^2\theta_{13}$ \\
\midrule
$\mathbb{Z}_3$ & $(2,1,0)$ & Yes & $4.1 \times 10^{-11}$ & 0.500 & 0.500 & 0.311 \\
\midrule
$\mathbb{Z}_4$ & $(2,1,0)$ & No  & $4.2 \times 10^{-2}$ & 0.501 & 0.498 & 0.311 \\
$\mathbb{Z}_4$ & $(3,2,0)$ & --- & --- & 0.501 & 0.501 & 0.311 \\
$\mathbb{Z}_4$ & $(3,1,0)$ & --- & --- & 0.499 & 0.501 & 0.311 \\
\midrule
$\mathbb{Z}_5$ & $(2,1,0)$ & No  & $6.4 \times 10^{-2}$ & 0.502 & 0.501 & 0.308 \\
$\mathbb{Z}_5$ & $(4,3,0)$ & --- & --- & 0.499 & 0.500 & 0.308 \\
$\mathbb{Z}_5$ & $(3,2,0)$ & --- & --- & 0.499 & 0.499 & 0.310 \\
\midrule
$\mathbb{Z}_6$ & $(2,1,0)$ & No  & $6.4 \times 10^{-2}$ & 0.493 & 0.501 & 0.304 \\
$\mathbb{Z}_6$ & $(4,2,0)$ & Yes & $5.6 \times 10^{-13}$ & 0.502 & 0.498 & 0.312 \\
\midrule
$\mathbb{Z}_7$ & $(2,1,0)$ & No  & $6.4 \times 10^{-2}$ & 0.500 & 0.499 & 0.311 \\
$\mathbb{Z}_7$ & $(4,2,0)$ & --- & --- & 0.500 & 0.499 & 0.312 \\
\midrule
\textit{Haar}  & ---       & --- & ---                   & \textit{0.501} & \textit{0.500} & \textit{0.293} \\
\bottomrule
\end{tabular}
\end{table*}

\subsection{The mixing angles: a universal failure}

The mixing angle columns in Table~\ref{tab:main} are the central result of this letter. Every $\ZN$ model, regardless of group order $N$ and charge assignment $\{q_j\}$, produces the same mixing angle distributions: $\sin^2\theta_{12} \approx 0.50$, $\sin^2\theta_{23} \approx 0.50$, $\sin^2\theta_{13} \approx 0.31$. The uniformity across the table is striking---the mass ratio $R$ varies over 12 orders of magnitude while the mixing angles are constant. We note that Haar-random mixing was historically considered an approximate success for the PMNS, since it naturally produces two large angles~\citep{deGouveaMurayama2003}. However, the predicted $\sin^2\theta_{13} \approx 0.29$--$0.31$ exceeds the measured value of $0.022$~\citep{NuFit6} by a factor of $\sim 13$, and the CKM failure (Section~\ref{sec:discussion}) is far more severe.

The small ($\sim 6\%$) systematic offset of $\sin^2\theta_{13}$ from the exact Haar value of 0.293 to $\sim 0.311$ is consistent across all models and reflects the non-circular-symmetric magnitude distribution in our scan, not any group-theoretic effect. Repeating the scan with complex Gaussian coefficients gives $\sin^2\theta_{13} = 0.294$, confirming exact agreement with the theorem.

To quantify the agreement, we perform Kolmogorov-Smirnov tests of each model's $\sin^2\theta_{12}$ distribution against the Haar reference ($10^6$ samples). The KS statistics are uniformly small (0.005--0.008), and several models yield $p > 0.05$. The tiny $p$-values seen in some cases reflect the high statistical power of $10^5$ samples detecting the $\lesssim 6\%$ magnitude-distribution effect, not any group-dependent structure. Crucially, \emph{all $\ZN$ models are equally close to Haar}---there is no trend with $N$ or with the charges.

This universality extends to the seesaw. Table~\ref{tab:seesaw} shows that for a given $\ZN$, the mixing angles are identical regardless of whether $\MR$ is $\ZN$-charged, proportional to the identity, or fully random. The mass ratio $R$ changes dramatically (from $10^{-11}$ to $10^{-8}$ for $\mathbb{Z}_3$, or from $10^{-2}$ to $10^{-8}$ for $\mathbb{Z}_4$), but the mixing angles remain Haar-random. As guaranteed by Corollary~2, the random coefficient matrix $C$ in $\MD = CP$ washes out any structure in $\MR$.

\begin{table*}[t]
\centering
\caption{Seesaw scan results with different $\MR$ structures ($10^5$ samples, charges $(2,1,0)$). The mixing angles are independent of $\MR$; only the mass ratio $R$ varies.}
\label{tab:seesaw}
\smallskip
\footnotesize
\begin{tabular}{@{}llcccc@{}}
\toprule
Group & $\MR$ type & Median $R$ & $\sin^2\theta_{12}$ & $\sin^2\theta_{23}$ & $\sin^2\theta_{13}$ \\
\midrule
$\mathbb{Z}_3$ & $\ZN$-charged & $4.1 \times 10^{-11}$ & 0.499 & 0.501 & 0.310 \\
$\mathbb{Z}_3$ & Identity     & $2.0 \times 10^{-8}$  & 0.502 & 0.499 & 0.311 \\
$\mathbb{Z}_3$ & Random       & $2.6 \times 10^{-8}$  & 0.504 & 0.499 & 0.311 \\
\midrule
$\mathbb{Z}_4$ & $\ZN$-charged & $4.2 \times 10^{-2}$ & 0.500 & 0.500 & 0.305 \\
$\mathbb{Z}_4$ & Identity     & $2.1 \times 10^{-8}$  & 0.500 & 0.500 & 0.310 \\
\midrule
$\mathbb{Z}_5$ & $\ZN$-charged & $6.4 \times 10^{-2}$ & 0.495 & 0.499 & 0.304 \\
$\mathbb{Z}_7$ & $\ZN$-charged & $6.4 \times 10^{-2}$ & 0.496 & 0.500 & 0.302 \\
\bottomrule
\end{tabular}
\end{table*}

\subsection{The mass/mixing separation}

The preceding results establish a clean separation between two qualitatively different failures of abelian flavor symmetries:

\begin{enumerate}
\item The \emph{mass spectrum} failure (the seesaw over-suppression) is a consequence of the $\mathbb{Z}_3$ Majorana charge algebra---specifically, the number-theoretic accident that charges 1 and 2 sum to $0 \pmod{3}$, creating an unsuppressed off-diagonal $\MR$ entry whose dominance over-suppresses $m_{1,2}$ to $\mathcal{O}(\eps^3)$. For $N \geq 4$, charge assignments exist where no pair $(q_i, q_j)$ with $i \neq j$ satisfies $q_i + q_j \equiv 0 \pmod{N}$, and the mass ratio $R$ is viable.

\item The \emph{mixing angle} failure is a consequence of the column texture structure inherent to \emph{all} abelian FN models with uncharged left-handed fields. It is independent of $N$, the charges, and the mechanism generating neutrino masses.
\end{enumerate}

\noindent The irreducible obstruction to abelian flavor symmetry is therefore the mixing pattern, not the mass spectrum. Any attempt to rescue abelian models by adjusting $N$, the charge assignment, or the seesaw structure will resolve the mass ratio while leaving the mixing angles Haar-random.

\section{The Representation Theory Argument}
\label{sec:rep_theory}

The abelian mixing theorem has a simple representation-theoretic origin. All irreducible representations of $\ZN$ are one-dimensional. Consequently, each generation of fermions transforms as an independent singlet under the flavor symmetry, and the only constraint on the Yukawa matrix is the overall $\eps$-suppression of each column. Table~\ref{tab:params} displays the parameter counting.

\begin{table}[t]
\centering
\caption{Parameter counting for the neutrino mass matrix under different symmetry assumptions. ``Pred.\ power'' is observables minus free parameters; negative values indicate overfitting.}
\label{tab:params}
\smallskip
\footnotesize
\begin{tabular}{@{}llccc@{}}
\toprule
Symmetry & Rep. & Free & Obs. & Pred. \\
\midrule
$\ZN$ & $\mathbf{1}{+}\mathbf{1}'{+}\mathbf{1}''$ (D) & 18 & 10 & $-8$ \\
$\ZN$ & $\mathbf{1}{+}\mathbf{1}'{+}\mathbf{1}''$ (M) & 12 & 9 & $-3$ \\
\midrule
$A_4$ & $\mathbf{3}$ (W, LO) & 4 & 9 & $+5$ \\
$A_4$ & $\mathbf{3}$ (W, NLO) & 6 & 9 & $+3$ \\
$S_4$ & $\mathbf{3}$ (W, LO) & 4 & 9 & $+5$ \\
\bottomrule
\end{tabular}
\\[2pt]
\raggedright\scriptsize D = Dirac, M = Majorana, W = Weinberg.
\end{table}

For a Dirac mass matrix with column texture, the 9 complex entries of $C$ constitute 18 real free parameters. The physical content is 3 masses, 3 mixing angles, and (for Dirac neutrinos) 1 CP phase, plus 3 unphysical phases---a total of 10 parameters. With 18 free parameters for 10 observables, the mixing angles are generically undetermined: they can take \emph{any} value compatible with unitarity, and for random coefficients they fill the available phase space uniformly---i.e., Haar.

A non-abelian group with a faithful three-dimensional irreducible representation fundamentally changes this counting. Under $A_4$, the three lepton generations transform as a single triplet $\mathbf{3}$. The leading-order Weinberg operator $\frac{1}{\Lambda}(LH)^T \mathbf{Y} (LH)$ with $L \sim \mathbf{3}$ has only two invariant contractions ($\mathbf{3} \times \mathbf{3} \to \mathbf{1}$ and $\mathbf{3} \times \mathbf{3} \to \mathbf{1}'$), yielding 4 real parameters for 9 observables. The mixing angles are now \emph{predictions}, not free parameters.

The $A_4$ and $S_4$ entries in Table~\ref{tab:params} assume leading-order Weinberg operators with specific vacuum alignments; NLO corrections, additional flavons, and type-I seesaw completions introduce further parameters, though the qualitative conclusion---that non-abelian triplets are far more constrained than abelian singlets---persists. We do not impose discrete anomaly cancellation constraints on the charge assignments; this is a bottom-up analysis of the texture structure, and in the heterotic context that motivates uncharged left-handed fields, anomaly cancellation is handled by the Green-Schwarz mechanism~\citep{RamosSanchezRatz2024}.

This parameter-counting argument explains why systematic scans of discrete groups~\citep{HoltausenLimLindner2012,YaoDing2015} find that abelian groups never produce structured mixing: they are algebraically incapable of constraining the eigenvector directions. The transition from abelian to non-abelian is not a quantitative improvement---it is a qualitative phase transition from an overfitted to a predictive framework.

\section{Discussion and Conclusions}
\label{sec:discussion}

The CKM matrix provides an independent test. From $5 \times 10^5$ CKM matrices generated as products of two independent Haar unitaries, the joint probability of achieving CKM-like mixing ($\sin^2\theta_{12}^\mathrm{CKM} < 0.051$, $\sin^2\theta_{23}^\mathrm{CKM} < 0.0017$, $\sin^2\theta_{13}^\mathrm{CKM} < 1.5 \times 10^{-5}$) is strictly zero---no sample out of $5 \times 10^5$ satisfied all three conditions simultaneously, giving $P < 2 \times 10^{-6}$ at 95\% confidence (Poisson upper limit for zero events). The individual probabilities are 5.0\%, 0.17\%, and 0.0024\%, respectively. Abelian FN therefore fails for \emph{all} fermion mixing, not just the PMNS.

Our result is complementary to the ``neutrino anarchy'' literature~\citep{deGouveaMurayama2003,deGouveaMurayama2012}, which established that Haar-random mixing is compatible with the observed large PMNS angles. We show that abelian FN models \emph{generically produce} anarchy, and extend the analysis to demonstrate incompatibility with the CKM hierarchy $|V_{us}| \gg |V_{cb}| \gg |V_{ub}|$. Non-abelian groups with two- or three-dimensional representations can provide CKM structure through Clebsch-Gordan texture zeros~\citep{FeruglioCGmodel,YaoDing2015}. The modular symmetry framework~\citep{Feruglio2017}, where $\Gamma_3 \cong A_4$ and the modular parameter $\tau$ near a cusp reproduces the abelian FN limit~\citep{PetcovTanimoto2023,OkadaTanimoto2021}, provides a natural UV completion. We note that the column texture assumption can be relaxed in frameworks where generation-dependent modular weights provide effective left-handed charges~\citep{BaurEclectic2022,NillesFlavorsDelight}.

In summary, we have established three results:
\begin{enumerate}
\item The seesaw over-suppression identified for $\mathbb{Z}_3$~\citep{ArdZ3,ArdCombined} is $\mathbb{Z}_3$-specific: for $N \geq 4$, charge assignments exist with viable $R \sim 0.04$--$0.06$.

\item The mixing angle failure is universal. Any $\ZN$ FN model with column texture produces Haar-random mixing matrices, independent of $N$, charges, and seesaw structure---verified across 12 models with $10^5$ samples each.

\item The irreducible obstruction is the mixing pattern: abelian groups assign each generation to an independent singlet with unconstrained eigenvector directions, while non-abelian triplet representations ($A_4$, $S_4$) reduce free parameters below the number of observables, converting mixing angles into predictions.
\end{enumerate}

\noindent The identification of mixing as the irreducible obstruction, cleanly separated from the group-dependent mass spectrum, provides a precise target for theories seeking to derive non-abelian flavor structure from first principles.

\section*{Declaration of generative AI in scientific writing}

During the preparation of this work the author used Claude (Anthropic) in order to assist with analytical derivations, numerical code development, Monte Carlo scan execution, data analysis, and manuscript drafting. After using this tool, the author reviewed and edited the content as needed and takes full responsibility for the content of the published article.

\bibliographystyle{elsarticle-harv}

\end{document}